\documentclass[usegraphicx,usenatbib,onecolumn,doublespace]{mn2e}
\usepackage{amsmath}
\usepackage{amssymb}
\usepackage{bm}

\begin{document}

\title{Constraining primordial non-Gaussianity with CMB-21cm cross-correlations?}

\author[Tashiro, H. \& Ho, S. ]
{Hiroyuki Tashiro$^1$, and Shirley Ho$^{2,3}$\\
$^1$
Physics Department, Arizona State University, Tempe, Arizona 85287, USA\\
$^2$
Lawrence Berkeley National Laboratory, 1 Cyclotron Rd, MS 50R-5045, Berkeley, CA 94720, USA\\
$^3$Carnegie Mellon University, Department of Physics, 5000 Forbes Ave., Pittsburgh, PA 15213, USA
}
\date{\today}

\maketitle

\begin{abstract}
We investigate the effect of primordial non-Gaussianity on the
cross-correlation between the CMB anisotropies and the 21cm fluctuations
from the epoch of reionization.  We assume an analytic reionization
model and an ionization fraction with $f_{\rm NL}$ induced scale
dependent bias. We estimate the angular power spectrum
of the cross-correlation of the CMB and 21~cm.  
 In order to evaluate the detectability,
the signal-to-noise (S/N) ratio for only a single redshift slice is also calculated for current and future
observations, such as CMB observations by  {\it Planck} satellite and
21cm observations by {\it Omniscope}.
The existence of the $f_{\rm NL}$ increases the signal of the
 cross-correlation at large scales and the amplification does not
 depend on the reionization parameters in our reionization model.
 However the cosmic variance is significant on such scales and
 the S/N ratio is suppressed.
 The obtained S/N ratio
is 2.8 (2.4) for $f_{\rm NL}=10$ ($100$)  in our fiducial
reionization model. 
Our work suggests in the 
absence of significant foregrounds and systematics, the
auto-correlations of 21~cm is a better probe of $f_{\rm NL}$ than the
cross-correlations (as expected since it depends on $b^2$), while the
cross-correlations contain only one factor of $b$. 
Nevertheless, it is interesting to examine the cross-correlations
between 21~cm and CMB, as the signal-to-noise ratio is not negligible
and it is more likely we can rid ourselves of systematics and foregrounds that
are common to both CMB and 21~cm experiments than completely clean
21~cm of all of the possible foregrounds and systematics in large scales.

\end{abstract}

\section{Introduction}
The inflationary scenario is strongly supported by the statistical
nature of the density fluctuations revealed by recent cosmic microwave
background (CMB) observations strongly.  The
observed density fluctuations have an almost scale invariant spectrum
and nearly Gaussian statistics as it was predicted by inflation
\citep{komatsu-wmap-2011}.

More recently, the measurement of the degree of deviation form
Gaussianity has attracted significant attention as this would help
pinpointing the correct model among very many inflationary models.  For
example, the density fluctuations arise from simple slow-roll inflationary
models with a single scalar field is almost purely Gaussian
\citep{guth-pi-1982, starobinsky-1982,bardeen-steinhardt-1983}, and the
deviation from the Gaussianity would then be unobservably small
\citep{falk-rangarajan-1993, gangui-lucchin-1994}.  On the other hand,
several inflationary models such as a single field inflation with
non-canonical kinetic terms or some multi-field inflation models can
generate primordial non-Gaussianities large enough to be observed by
ongoing surveys, e.g. {\it Planck} \citep{planck-bluebook} (For
comprehensive review see \citealt{bartolo-komatsu-2004}).

Wilkinson Microwave Anisotropy Probe (WMAP) puts one of the strongest
constraints on 
the local type of the primordial non-Gaussianity
\citep{komatsu-wmap-2011},
which is parameterized by the constant dimensionless parameter
$f_{NL}$ as \citep{komatsu-spergel-2001}
\begin{equation}
\Phi({\bm x}) = \Phi_{\rm G}({\bm x}) +f_{\rm NL}(\Phi_{\rm G}^2({\bm x}) 
- \langle \Phi_{\rm G}^2 ({\bm x})\rangle), 
\end{equation}
where $\Phi$ is Bardeeen's gauge-invariant potential,
$\Phi_{\rm G}$ is the Gaussian part of the potential and
$\langle ~ \rangle$ denotes the ensemble average.
For example, the present constraints on the local type of $f_{\rm NL}$ from WMAP are 
$-18<f_{\rm NL}<80$ by \citet{curto-martinezgonzales-2009} and 
$-36<f_{\rm NL}<58$ by \citet{2009PhRvD..80l3005S}

The effect of primordial non-Gaussianity appears not only on CMB
fluctuations but also on large scale structure.  The abundance and
clustering of virialized objects is sensitive to the existence of the
primordial non-Gaussianity, as it was first discussed by
\citet{dalal-2008}, and \citet{slosar-2008} has shown that competitive
constraints can be achieved from large scale structure, $-29< f_{\rm NL}
< 70$.  High redshift galaxy survey with large volumes are also expected
to be good probes for the primordial non-Gaussianity
\citep{desjacques-seljak-2010}.

With the ongoing and upcoming surveys of 21-cm, such as LOFAR
\citep{2010MNRAS.405.2492H},
MWA
\citep{2009IEEEP..97.1497L},
SKA
\citep{2004NewAR..48..979C} and Omniscope \citep{2010PhRvD..82j3501T},
we will soon have a map of the 21-cm emission line of a large volume of
the Universe.  21~cm line emission comes from the spin-flip transition
of neutral hydrogen. 21~cm fluctuations depends on the abundance and
clustering of ionization sources, it is expected to reveal the EoR by
the observation of the 21~cm fluctuations.  Since the fluctuations of
the 21~cm line emission trace the large scale structure, they are also
expected to be sensitive for primordial non-Gaussianity.

The virialized objects such as first stars and galaxies are considered
as possible ionization sources, therefore the constraint on the
primordial non-Gaussianity can be obtained through the study on the
21~cm fluctuations induced by ionization sources evolved from initial
conditions with primordial non-Gaussianity.  For example,
\citet{2011arXiv1105.1773J} have investigated the 21~cm power spectrum
with $f_{\rm NL}$.
They showed that SKA and MWA could measure $f_{\rm NL}$
values of order 10 and Omniscope has the potential to put much
more stringent constraint $f_{\rm NL} \sim 1$.
\citet{2011arXiv1104.0149T} have studied
the bubble number count with $f_{\rm NL}$ on 21~cm maps.
With the imminent release of CMB maps from Planck and looking forward
to various 21~cm experiments,
in this paper, we set out to investigate the effect of primordial non-Gaussianity on
the cross-correlation between CMB temperature anisotropies and 21~cm
fluctuations from the epoch of reionization (EoR) in the analytic
reionization model.

The cross-correlation of CMB-21~cm is
expected to be powerful tools for investigating the evolution of the
cosmic reionization \citep{2006ApJ...647..840A, 2008MNRAS.384..291A}, and is expected to be
detected by near future observations \citep{2010MNRAS.402.2617T}.
As the removal of foreground and systematic effects in 21~cm is a
challenging task, it is beneficial to invest in another method other
than the auto-correlations of 21~cm. In particular, employing
cross-correlations with another independent experiment reduces the
possible extra-power in auto-correlations due to systematics or foregrounds within one
single experiment.
The cross-correlation amplitude strongly depends on the evolution of the
ionized fraction \citep{2006ApJ...647..840A}.  Since the existence of
the non-Gaussianity is also expected to amplify the amplitude, it is
important to evaluate the effect of primordial non-Gaussianity in order
to extract the information about the EoR from the cross-correlation
between CMB and 21~cm fluctuations.  In this paper, we focus on the
non-Gaussianity effect on the ionized fraction through calculating the
scale dependent bias of ionized bubbles due to the primordial
non-Gaussianity based on \citet{dalal-2008}, as ionized
bubbles are density peak tracers of the density fluctuations in the
EoR.  This particular
formalism is applicable not only for 21~cm
fluctuations but also for other large scale structures which are also
density peak tracers.  We also calculate the angular power spectrum of 21~cm
fluctuations with the primordial non-Gaussianity for comparison.

The outline of this paper is the following.  In Sec.~II, we present the
analytic representation for the cross-power angular spectrum of CMB
temperature anisotropies and 21~cm fluctuations.  In Sec.~III, we give
the simple analytic model of the EoR and the ionized fraction bias.  In
Sec.~IV, we show the results of the cross-power spectrum and discuss the
effect of primordial non-Gaussianity. Section V is devoted to the
conclusion.  Throughout the paper, we use the concordance cosmological
parameters for a flat cosmological model, i.e. $h=0.7 \ (H_0=h \times
100 {\rm ~km/s / Mpc})$, $\Omega _{\rm b} =0.05$, $\Omega_{\rm m} =0.26$
and $\sigma_8=0.8$.

\section{21cm--CMB cross-correlation during the epoch of reionization}

\subsection{Cosmological 21~cm signal during the EoR}
  
The observed differential brightness temperature of the
21~cm line from the redshift $z$ in the direction of $\hat {\bm n}$
is given as in \cite{madau-meiksin-rees-1997} by
\begin{equation}
T_{\rm B} (\hat {\bm n} , z) = T_{21}(z) \overline{ x}_{\rm H} (z)
(1+\delta_{x _{\rm H}}(\hat {\bm n} , z) )
\left(1+\delta(\hat {\bm n} , z) - {1 \over a(z) H(z)} {\partial v_r
 \over \partial
r} \right)  ,
\label{eq:21cmline}
\end{equation}
where $\overline{ x}_{\rm H}$ is the hydrogen neutral fraction,
$\delta_{x _{\rm H}}$ is the fluctuation contrast of $x_{\rm H}$,
$\delta$ is the baryon density contrast, and $d v_r/ dr$ is the gradient
of the radial velocity $v_r$ along the line of sight. The term including
$d v_r /dr $ accounts the redshift distortion due to the bulk motion of
the hydrogen \citep{kaiser-1987}.  
The normalization temperature $T_{21}(z)$ is written as
\begin{equation}
T_{21} = 26~ \overline{x}_{H}
\left(1-{T_{\rm CMB} \over T_{\rm s}} \right) 
\left({\Omega_b h^2 \over 0.02} \right) 
\left[ \left({1+z \over 10} \right)
\left({0.3 \over \Omega_{\rm m}} \right) \right]^{1/2}
~{\rm mK} ,
\end{equation}
where $T_{\rm CMB}$ is the CMB temperature and $T_{\rm s}$ is the spin
temperature given by the ratio of the number density of hydrogen in the
excited state to that of hydrogen in the ground state.  During the epoch
of reionization, the spin temperature becomes much larger than the CMB
temperature \citep{ciadri-madau-2003}.  Therefore we assume $(1-{T_{CMB}
/ T_s}) \sim 1 $ hereafter.

Applying the spherical harmonic transformation to
Eq.~(\ref{eq:21cmline}), we can obtain the multipole moments of the
21~cm fluctuations $a^{21}_{\ell m}$.
In the linear order,
$a^{21}_{\ell m}$ is expressed as
\begin{equation}
a^{21}_{\ell m} (z) = 4 \pi (-i)^\ell T_{21}(z)
D(z) \overline{x}_{\rm H} \int {d^3 k \over (2 \pi)^3}
\left( \delta_{k} J_\ell (k r_z) + \delta_{x_{\rm H} k} j_\ell (k r_z) \right)
Y^{*}_{\ell m} (k),
\label{eq:alm_21cm}
\end{equation}
where $r_z$ is the radial distance to the redshift $z$, $r_z= \eta_0 -\eta(z)$ 
with the conformal time at the present epoch $\eta_0$,
$j_\ell (x)$ is the spherical Bessel function and
\begin{equation}
 J_\ell(x) = -{\ell (\ell-1) \over 4 \ell^2 -1} j_{\ell-2} (x)
  +\left({2 \ell^2 +2 \ell -1 \over 4 \ell^2 +4 \ell -3} +1
   \right) j_\ell (x)
  -{(\ell+2 ) (\ell+1) \over (2\ell +1) (2\ell +3)]) } j_{\ell+2}(x),
\end{equation}
in the matter dominated epoch \citep{bharadwaj-ali-2004}.
In Eq.~(\ref{eq:alm_21cm}),
$D(z)$ is the linear growth factor, and $\delta_{k}$
and $\delta_{x_{\rm H} k}$ are 
the Fourier components of $\delta$
and $\delta_{x_{\rm H}}$, respectively.

\subsection{CMB Doppler signal from the EoR}

The main contribution of the CMB temperature anisotropy on large scales
from the EoR comes from the Doppler effect.  The CMB Doppler effect in
the direction ${\hat{\bm n}}$ is given in the linear order by
\begin{equation}
T_{\rm D}({ \hat{\bm n}})= -T_{\rm CMB}
\int_0^{\eta_0} d\eta \dot{\tau}e^{-\tau}{\hat{\bm n}}\cdot 
{\bm  v} ({\hat{\bm n}},\eta),
\label{eq:CMBdoppler}
\end{equation}
where ${\bm v}$
is the peculiar velocity of baryons, $\dot \tau$ is the differential
optical depth for Thomson scattering $\tau(\eta)$ in conformal time,
which is given by $\dot \tau = n_e \sigma_{\rm T}a$, with the electron
number density $n_e$, the scale factor $a$ and the cross section of
Thomson scattering $\sigma_{\rm T}$.  The continuity equation gives the
relation between the peculiar velocity and the density contrast.
\begin{equation}
{{\bm v}_{{\bm k}}}=-i({\bm k}/{k^2}) 
\dot \delta_{\bm k},
\label{eq:continuity}
\end{equation}
where the dot represents the derivative with respect to conformal time.

Using the spherical harmonics expansion to Eq.~(\ref{eq:CMBdoppler})
with Eq.~(\ref{eq:continuity}), we obtain the multipole components of
the CMB Doppler anisotropy in the linear order,
\begin{equation}
a_{\ell m} ^{\rm D} =  4 \pi (-i) ^\ell
\int d \eta \int {d^3 k \over (2 \pi)^3} T_{\rm CMB}
\dot D \dot \tau e^{-\tau} {\delta_{k} \over k^2}
{\partial \over \partial \eta}
j_\ell(k r) Y^{*}_{\ell m} (k),
\label{eq:alm_doppler}
\end{equation}
where $r=\eta_0-\eta$.

\subsection{Cross-correlation}\label{sec:cross-corr}

The angular power spectrum of the cross-correlation between the 21~cm
fluctuations from the redshift $z$ and CMB anisotropies $C_{\ell} (z)$
can be obtained from the ensemble average of both the multipole
components .  From Eqs.~(\ref{eq:alm_21cm}) and (\ref{eq:alm_doppler}),
$C_{\ell} (z)$ is expressed as \citep{2006ApJ...647..840A,2008MNRAS.384..291A},
\begin{equation}
C_\ell (z) =
-T_{\rm CMB} T_{21} (z) D(z) \frac{2}{\pi} 
\int_0^\infty dk
\int_0^{\eta_0} d\eta' 
\left[(1-\overline{x}_i(z))
P_{\delta}( k) J_\ell (k r_z)
- \overline{x}_i(z) P_{x \delta}  j_\ell (k r_z) \right]
j_\ell(k r')
\frac{\partial}{\partial \eta'}(\dot D \dot{\tau}e^{-\tau})  ,
\label{eq:cross-21T-0}
\end{equation}
where $r' = \eta_0 -\eta'$,
$P_{\delta}$ is the matter power spectrum, $P_{x \delta}$ is the
cross-power spectrum between $\delta$ and $\delta_x$.  In order to
obtain Eq.~(\ref{eq:cross-21T-0}), neglecting the effect of helium
ionization, we assume $\overline{ x}_i = 1- \overline{ x}_{\rm H}$ and
$\delta_x = -\delta_{x_{\rm H}}$.  Eq.~(\ref{eq:cross-21T-0}) tells us that
the amplitude of the cross-correlation strongly depends on the evolution
of the ionization fraction through $\partial \dot{\tau} / \partial
\eta$.



\section{The bias of ionized fraction fluctuations}

In order to calculate Eq.~(\ref{eq:cross-21T-0}), it is required to
evaluate $P_{x \delta}$ which depends on the reionization model. Since
we focus on large scales in this paper, we assume that $\delta_x$ can be
written with $\delta $ and the constant bias $b_x$ as
\begin{equation}
\delta_x = b_x \delta.
\label{eq:deltax-bias}
\end{equation}

The bias $b_x$ depends on the reionization model. In this paper,
we adopt the analytical model based on the `inside-out' reionization
scenario \citep{furlanetto-2006}. For simplicity, we do not include the
effect of the primordial non-Gaussianity in this section.

First, we assume that a galaxy of mass $m_{\rm gal}$ can
ionize a mass $\zeta m_{\rm gal}$ where $\zeta$ is the efficiency factor
for the reionization process. Therefore, the background ionized fraction
can be associated to the collapsed fraction $f_{\rm coll}$ which is the
fraction of mass in halos above the mass threshold for collapse,
$m_{\rm min}$ \citep{furlanetto-2006},
\begin{equation}
 {d \bar {x}_i \over d t} = \zeta {d f_{\rm coll} \over dt }
  - \alpha \bar{x}_i n_e (z) C,
\end{equation}
where $\alpha$ is the recombination coefficient at $10^4$ K, $\alpha =
4.2 \times 10^{-12} ~\rm cm^{3} s^{-1}$, $C$ is the clumping factor for
ionized gas. 

Now we divide space into cells of mass $m$.  The different cells have
different density fluctuations $\delta$. Since we assume that a galaxy
with mass $m_{\rm gal}$ can ionize mass $\zeta ma_{\rm gal}$, the
ionized fraction in the cell of mass $m$ with the density fluctuations
$\delta$ can be written as
\begin{equation}
x_i(\delta) = \bar{x}_i (1+\delta_x) =
 - \zeta f_{\rm coll}(m, \delta),
\label{eq:def-xi}
\end{equation}
where $\zeta $ is $f_{\rm coll}(m, \delta)$ is the conditional collapse
fraction in the cell of mass $m$ with $\delta$.
According to the extended Press-Schechter theory, $f_{\rm
coll} (m, \delta)$ is given by
\begin{equation}
f_{\rm coll} (m, \delta) = {\rm erfc} \left[ 
{\delta_c (z)- \delta (z)
\over \sqrt {2 (\sigma_{\rm min}^2 -\sigma^2(m))}} 
\right],
\label{eq:fcoll-delta}
\end{equation}
where $\delta _c(z)$ is the critical density for collapse at the redshift $z$, $\sigma(m)$ is
the smoothed dispersion of the initial density fields with the top-hat window
function associated with mass $m$ and $\sigma_{\rm min}$ is the smoothed
dispersion at the mass $m_{\rm min}$.

Since we are interested in large scales, we consider cells associated
with large mass $m$.  The rms density fluctuation in such a cell is much
smaller than unity. Therefore, we can assume that $|\delta |\ll1$ and
$\sigma(m) \ll1 $ in most of cells.  Applying the Taylor series
expansion to Eq.~(\ref{eq:fcoll-delta}), we
obtain
\begin{equation}
f_{\rm coll} \approx f_{\rm coll}(0)[1+\overline{ b}  \delta],
\label{eq:fcoll_delta}
\end{equation}
where
\begin{equation}
\overline{ b} \equiv \sqrt{2 \over \pi} {\exp[-\delta_c^2 (z)
/2 \sigma^2_{\rm min} ] \over f_{\rm coll} \sigma_{\rm min} D(z)}.
\end{equation}

\begin{figure}
  \begin{center}
\includegraphics[keepaspectratio=true,height=60mm]{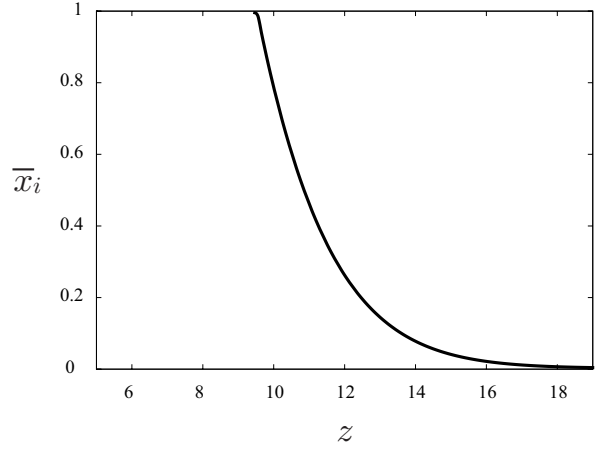}
  \end{center}
  \caption{The redshift evolution of the ionized fraction. 
we adopt $\zeta=40$ and $m_{\rm min}$ corresponding
to the virial temperature $T_{\rm vir}=10^4~$K.}
  \label{fig:ionization-history}
\end{figure}

According to Eqs.~(\ref{eq:deltax-bias}), (\ref{eq:def-xi}) and
(\ref{eq:fcoll_delta}), we take $b_x = \bar b $ in our model.
Our reionization model has three parameters, $\zeta$, $m_{\rm min}$ and $C$.
In this paper, we adopt $\zeta=40$ and $m_{\rm min}$ corresponding
to the virial temperature $T_{\rm vir}=10^4~$K \citep{2001PhR...349..125B},
\begin{equation}
m_{\rm min} =  3.3 \times 10^7 \left(\frac{T_{\rm vir}}{10^4}\right)^{3/2}
 \left(\frac{z+1}{10}\right)^{-3/2} \left(\frac{h^2 \Omega_{\rm
  m}}{0.147}\right)^{-1/2} M_\odot.
\end{equation}
Although $C$ is calculated with numerical simulations and
depends on how the IGM was ionized, we set $C$ in order to the
ionized fraction reach 0.5 at $z=11$ for simplicity.
Figure~\ref{fig:ionization-history} shows the evolution of
the mean ionized fraction.

\section{Effects of the primordial non-Gaussianity}

Primordial non-Gaussianities modify the abundance, merger history and
clustering of dark halos \citep{dalal-2008,slosar-2008}. These effects
will evidently induce the
early beginning of the reionization process, even though the effect on the
reionization optical depth is not very strong. For example, the local type
non-Gaussianity with $f_{\rm NL} =100$ will only enhance the optical depth by $1\%$
\citep{2009MNRAS.394..133C}.  Therefore we assume that the redshift
evolution of the background ionized fraction obtained in the previous
section is not modified by the existence of the non-Gaussianity.
However, as reionization sources are collapsed objects in the
fluctuation of the density field, the fluctuations of the ionized fraction can be affected strongly
by primordial non-Gaussianity. 

\citet{dalal-2008}
have studied the effect of $f_{\rm NL}$ on
peak heights of the density fluctuations and the bias of density
tracers.  According to Eq.~(9) in \citet{dalal-2008},
we can
write the bias of density fluctuations of ionized fraction with
primordial non-Gaussianity is given by
\begin{equation}
b_{\rm NG} = b_x +2(b_x-1) f_{\rm NL} \delta_B {3 \Omega_m H_0^2 \over 
2 a D(a) k^2 T(k)},
\label{eq:def-bng}
\end{equation}
where T(k) is the transfer function ($T(k)\sim 1$ on large scales),
$\delta_B$ is the critical density for the bias tracer. In this paper,
the bias tracer is the ionized cell. Therefore we adopt the critical
density for the ionized cell, while the critical density is for
collapse, $\delta_c$, in \citet{dalal-2008}.

Since the ionized fraction is almost unity in an ionized cell, $\zeta
f_{\rm coll} \gg 1$ in such a cell. Therefore we assume that the
condition for the ionized cell is $\zeta f_{\rm coll} > 1$.  This
condition gives the critical density for the ionized cell,
\begin{equation}
\delta_B \equiv \delta_c - \sqrt{2} K(\zeta)
[\sigma_{\rm min}^2 - \sigma_2 (m)]^{1/2},
\label{eq:deltab-def}
\end{equation}
where $K(\zeta) ={\rm erf} ^{-1} (1-\zeta^{-1})$.
Eq.~(\ref{eq:deltab-def}) corresponds to Eq.~(4) in
\citet{furlanetto-zaldarriaga-2004} and Eq.~(5) in
\citet{2011arXiv1105.1773J}.
The primordial non-Gaussianity $f_{\rm
NL}$ affect $f_{\rm coll} (\delta)$.  However this effect is enough
small that we can neglect the modification of $f_{\rm NL}$ on $f_{\rm
coll} (\delta)$ ( \citealt{2011arXiv1105.1773J} reported that the mean
critical density $\overline{\delta}_B$ is only perturbed by 4\% even for
$f_{\rm NL}=100$).  Now we can write the cross-power spectrum between
$\delta$ and $\delta_x$ with $f_{\rm NL}$ is given by
\begin{equation}
P_{\delta x} (k)=b_{\rm NG} P_{\delta}(k).
\label{eq:cross-fnl}
\end{equation}

\begin{figure}
  \begin{center}
\includegraphics[keepaspectratio=true,height=60mm]{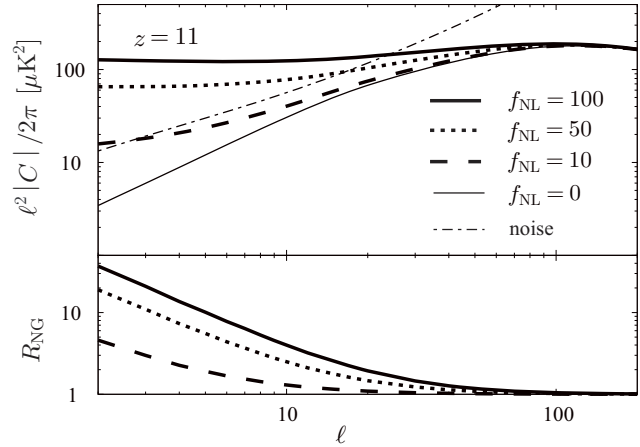}
  \end{center}
  \caption{The cross-power spectrum for different $f_{\rm NL}$ 
(top panel) and the ratio of the cross-power spectrum between
non-Gaussian and Gaussian cases (bottom panel).  On both panels,
the solid, dotted and dashed lines are for $f_{\rm NL} =100$,
$f_{\rm NL} =50$ and $f_{\rm NL} =10$, respectively. For comparison, we 
plot the angular power spectrum for the Gaussian case as
the thin solid line on the top panel.
The dashed-dotted line represents the noise power spectrum in the
 combination of {\it Planck} and Omniscope.
 }
  \label{fig:cross-angular}
\end{figure}

Using Eqs.~(\ref{eq:cross-21T-0}) and (\ref{eq:cross-fnl}),
we calculate the cross-power spectra between CMB and the 21~cm
line from $z=11$ for different $f_{\rm NL}$s.
We plot the angular power spectrum of the
cross-correlation in the top panel of Figure~\ref{fig:cross-angular}.
We also show the ratio of the angular spectra 
between non-Gaussian and Gaussian cases, $R_{\rm NG} =
C^{\rm NG}_{\ell}/C^{\rm Gaussian}_{\ell}$, in the bottom panel.
Due to the scale dependent bias introduced by $f_{\rm NL}$,
the higher $f_{\rm NL}$  induces higher cross-correlation on large
scales, while the effect of $f_{\rm NL}$ on the cross-correlation is small on smaller 
scales ($\ell >100$)  and it
does not modify the position and the height of the 
cross-correlation peak.
As pointed by \citet{2006ApJ...647..840A, 2008MNRAS.384..291A}, the peak height of the cross-correlation
depends on the evolution of the ionized fraction.
Therefore, these facts suggest that the spectrum of the 
cross-correlation on large scale have the potential to give 
the constraint on $f_{\rm NL}$, while one can derive information on
the evolution of the cosmic reionization from the  peak height and position.

\begin{figure}
  \begin{center}
\includegraphics[keepaspectratio=true,height=60mm]{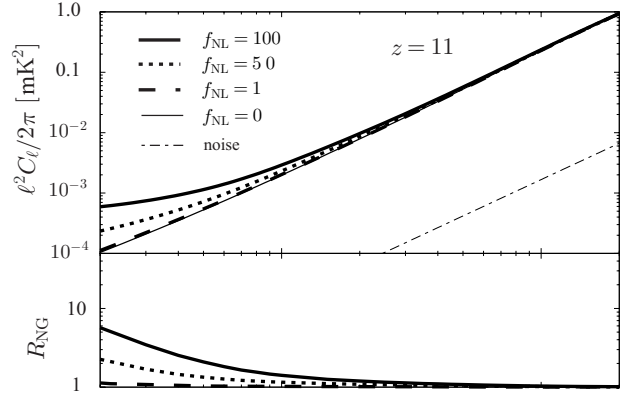}
  \end{center}
  \caption{The angular spectrum of 21~cm fluctuations for 
different $f_{\rm NL}$ 
(top panel) and the ratio of the angular spectrum spectrum between
non-Gaussian and Gaussian cases (bottom panel). We adopt the fiducial
reionization model ($\zeta=40$, $T_{\rm vir}=10^4~$ K). 
Types of lines
are same as in Figure~\ref{fig:cross-angular}.
The dashed-dotted line represents the noise power spectrum of Omniscope.
 }
  \label{fig:21cm-angular}
\end{figure}

For the purpose of cross-checks and comparisons, we also calculate the angular power spectrum of
21~cm fluctuations $C_\ell ^{21}$.
According to Eq.~(\ref{eq:alm_21cm}), $C_\ell ^{21}$ can be 
written as
\begin{equation}
C_\ell^{21} (z) =
T_{21} (z)^2  
\frac{2}{\pi} \int_0^\infty dk ~k^2
\left[(1-\overline{x}_i(z))^2
P_{\delta}( k) J_\ell ^2 (kr)
- 2 \overline{x}_i(z)  (1-\overline{x}_i(z)) P_{x \delta} J_\ell (kr) j_\ell(kr) 
+\overline{x}_i(z)  P_{x x}  j_\ell ^2(kr)\right],
\label{eq:spectrum-21}
\end{equation}
where $P_{xx}$ is the power spectrum of the ionized fraction
and is assumed to be $P_{xx}(k)=b_{\rm NG}^2 P(k)$.
Note that we have neglected the redshift distortion here due to the 
peculiar velocity of baryons. 
The top panels of Figure~\ref{fig:21cm-angular} shows $C_\ell^{21}$ 
at $z=11$ for different $f_{\rm NL}$ and the bottom panel represents
the ratio of the angular spectra between non-Gaussian and Gaussian
cases, $R_{\rm NG}$.
The angular spectrum on large scales is also amplified by non-zero 
$f_{\rm NL}$, showing similar behavior in  the 21 cm power spectrum as
discussed in 
\citet{2011arXiv1105.1773J}.
Compared with the cross-correlation,
auto-correlation has a larger amplitude, but the degree of the amplification due to $f_{\rm NL}$ is small on scales
$\ell < 10$. This is because the term $P_{xx}$ proportional to $b_{\rm
NG}^2$ is partially canceled by $P_{x \delta}$ on these scales.

\begin{figure}
  \begin{center}
\includegraphics[keepaspectratio=true,height=60mm]{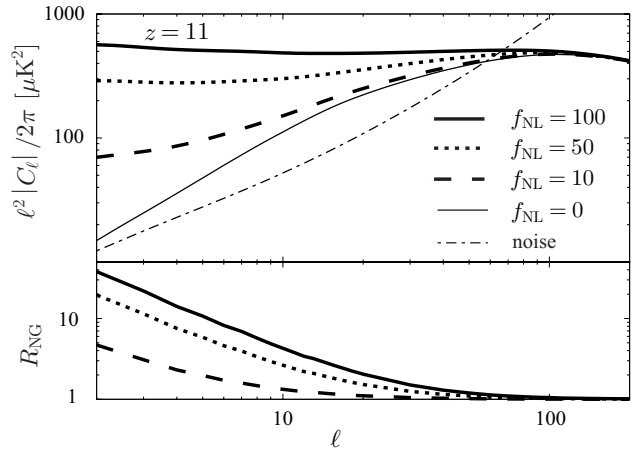}
  \end{center}
  \caption{The cross-power spectrum in the {\it rapid} reionization
    model for different $f_{\rm NL}$ 
(top panel) and the ratio of the angular spectrum spectrum between
non-Gaussian and Gaussian cases (bottom panel). Types of lines
are same as in Figure~\ref{fig:cross-angular}.The dashed-dotted line represents the noise power spectrum in the
 combination of {\it Planck} and Omniscope.}
  \label{fig:cross-angular-high}
\end{figure}

On the other hand, as we can see from Eq.~(\ref{eq:cross-21T-0}), the
cross-correlation amplitude includes $\partial \dot{\tau} /\partial
\eta$, the amplitude of the cross-correlation depends on the efficiency
of the reionization process.  We evaluate the cross-correlation in the
following {\it rapid} reionization model with $\zeta=400$ and $T_{\rm
vir}=10^5~$K. These parameters are motivated by the scenario in which
the sources of ionization photons are massive objects like QSOs.  We
plot the angular power spectrum of the cross-correlation for the {\it
rapid} reionization model in the top panel and the ratio of the angular
spectra between non-Gaussian and Gaussian cases in of
Figure~\ref{fig:cross-angular-high}.

As discussed in \citet{2006ApJ...647..840A, 2008MNRAS.384..291A}, the
height of the peak is amplified by the rapidness of the reionization
process.
The amplification due to $f_{\rm NL}$ depends on $\delta_B$.
According to Eq.~(\ref{eq:deltab-def}), $\delta_B$ is related to the reionization
parameters, $\zeta$ and $T_{\rm vir}$.
Large values of $\zeta$ increase $\delta_B$, while large values of $T_{\rm vir}$
make $\delta_B$ small  through decreasing $\sigma_{\rm min}$.
As a result, $\delta_B$ in the {\it rapid} reionization model is almost
same as in the fiducial model. Therefore, the dependence of $R_{\rm NG}
$ on $f_{\rm NL}$ in the bottom panel of figure~\ref{fig:cross-angular-high}
is same as in the fiducial model shown in figure~\ref{fig:cross-angular}.

\subsection{The Detectability of the Cross-Correlation}

In this section, we calculate the signal-to-noise (S/N) ratio of the cross-correlations
in order to study the detectability of the cross-spectrum 
signal with $f_{\rm NL}$.

If we assume that foreground correlation between CMB and 21~cm
can be removed \citep{2006ApJ...653..815M, 2006ApJ...648..767M},
the S/N ratio for the cross-correlation can then be expressed as
\begin{equation}
\left( {S \over N} \right) ^2 =
f_{\rm sky} \sum_{\ell = \ell_{\rm min}} ^{\ell_{\rm max}} (2 \ell +1)
{| C_\ell ^{21-\alpha}| ^2  \over | C_\ell ^{21-\alpha} |^2
+(C_\ell ^{21} +N_\ell ^{21} ) (C_\ell ^{\bm CMB} +N_\ell ^{\bm CMB})}.
\label{eq:SNratio}
\end{equation}
where $f_{\rm sky}$ is the sky fraction
common to both CMB and 21~cm observations,
the superscript $21$ stands for 21-cm fluctuations,
the superscript $CMB$ stands for the CMB anisotropy,
and $C_\ell$ and $N_\ell$ are the signal and 
the noise power spectrum, respectively.
According to Eq.~(\ref{eq:SNratio}), even if there is no instrumental noise ($N_\ell =0$),
the S/N ratio is limited by the cosmic variance, especially for small
values of $\ell_{\rm max}$.
In this paper, since we are interested in large scales, we 
consider {\it Planck}
as CMB observation whose sky fraction is 
almost unity. Therefore $f_{\rm sky}$  corresponds to the one
of the considered 21~cm observation.
In the {\it Planck} configuration, compared with the CMB signal,
the experimental noise is very small on our scales of interest,
$C_\ell^{\rm CMB} \gg N_\ell^{\rm CMB}$.
Therefore, we neglect $N_\ell^{\rm CMB}$ in the calculation.

The noise power spectrum of 21-cm observations is given by
\begin{equation}
{ N_\ell ^{21}}= 
 { 2 \pi \over t_{\rm obs} \Delta \nu} \left( { \ell_{\rm
max} \over 2 \pi} {\lambda^2 T_{\rm sys}\over A_{\rm eff}} 
\right)^2. 
\end{equation}
where $\Delta \nu$ is the bandwidth, $t_{\rm  obs}$ is the total
observation time, and $\ell_{\rm max}=
2 \pi {D / \lambda}$ is the maximum multipole associated with the
length of the baseline $D$. 
The system temperature $T_{\rm sys}$ is dominated by sky temperature
which is expressed as $T_{\rm sys} =2.7 (1+z)^{2.3} ~\rm K$ 
\citep{2006ApJ...638...20B}.
$A_{\rm total}$ is the total effective area which is assumed to be
$A_{\rm total} = N A_{\rm eff}$ with N being the number of the antenna
and  $A_{\rm eff}$ being the effective area of one antenna.
For 21~cm observation, we consider an optimistic experiment Omniscope (or FFTT) 
\citep{2009PhRvD..79h3530T}. We evaluate the
observational noise with $N=10^6$, $A_{\rm eff}=1~ {\rm m}^2$, 
$D=1~$km, $t_{\rm obs}=4000~$hours and $f_{\rm sky}=2 \pi$
\citep{2008PhRvD..78b3529M}. 
We plot the estimated noise power spectrum 
as the dashed-dotted lines in Fig.~\ref{fig:cross-angular} and \ref{fig:cross-angular-high}.

We also show the noise power spectrum of Omniscope in the
auto-correlation of 21~cm fluctuations in Fig.~\ref{fig:21cm-angular}.
Compared with the auto-correlation, the noise of the cross-correlation
is large in particular, on small scales, because CMB Doppler signal from
the EoR is proportional to $k^{-1}$ and the primordial CMB signal gives
large noise on small scales.

\begin{figure}
 \begin{minipage}{0.5\hsize}
  \begin{center}
   \includegraphics[keepaspectratio=true,height=55mm]{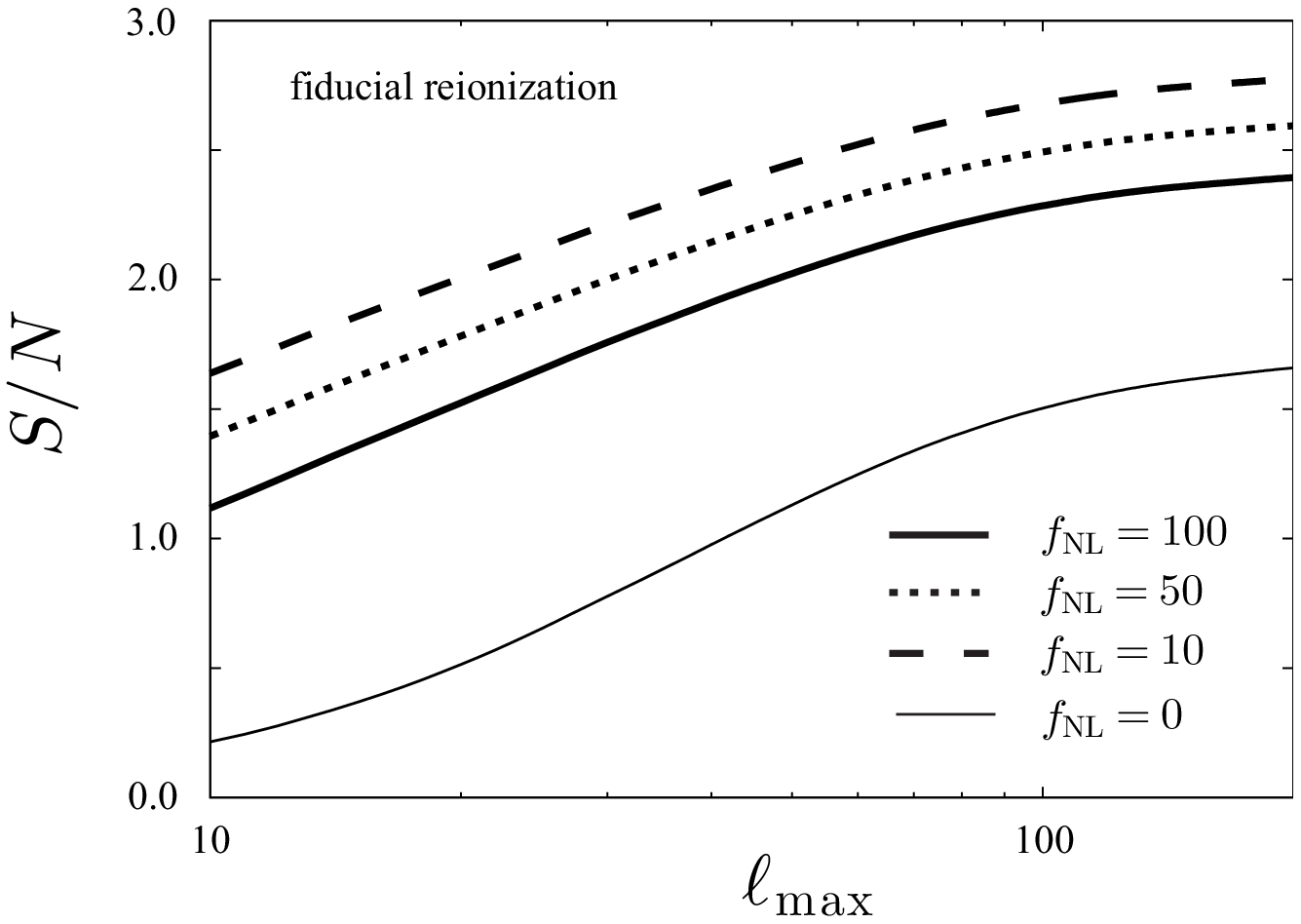}
  \end{center}
 \end{minipage}
 \begin{minipage}{0.5\hsize}
  \begin{center}
      \includegraphics[keepaspectratio=true,height=55mm]{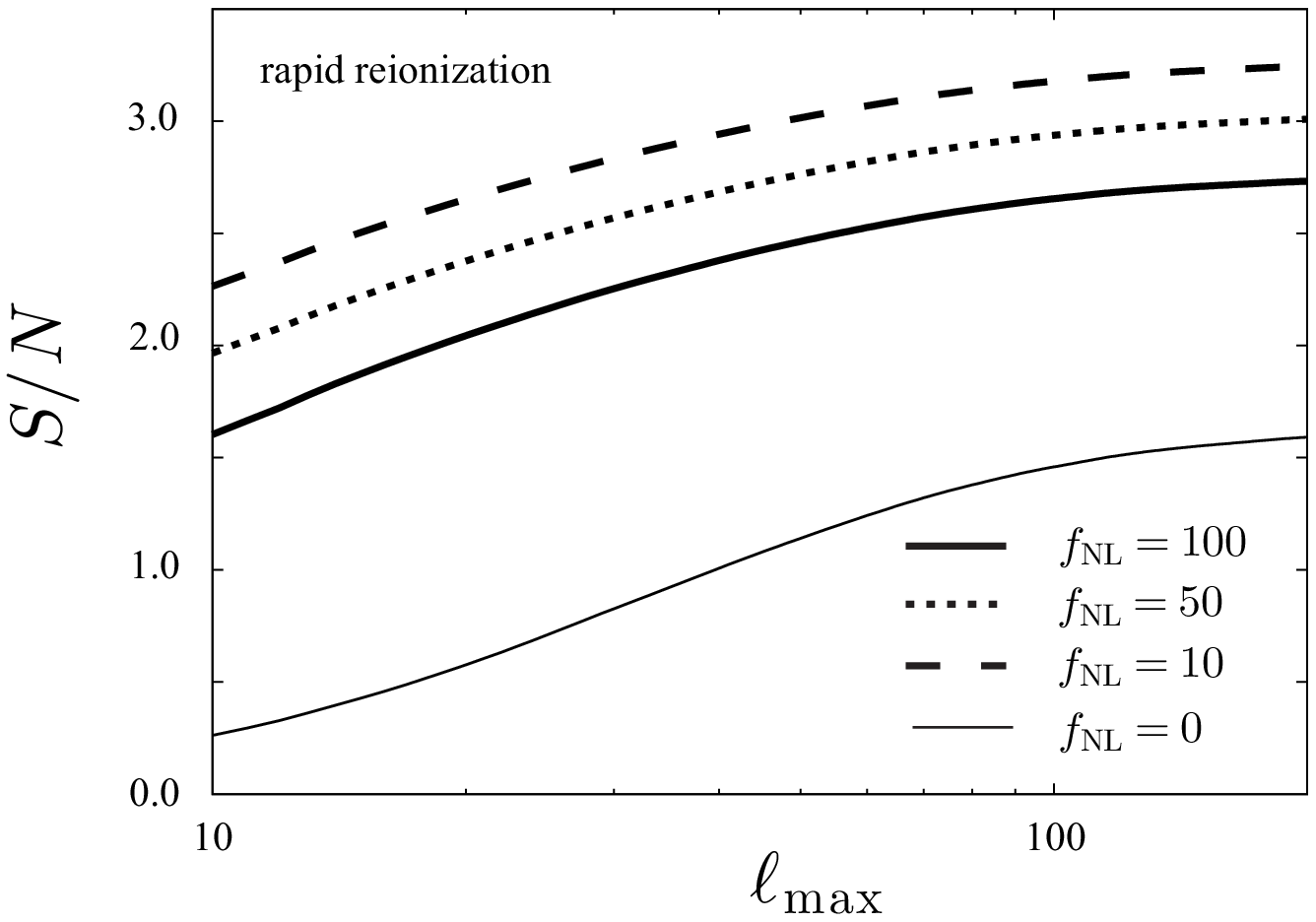}
  \end{center}
 \end{minipage}
 \caption{
 The dependence of the S/N ratio on $\ell_{\rm max}$ for the
 combination of {\it Planck} and Omniscope.
The left and right panels show
for the fiducial (slow)  and {\it rapid} reionization model,
 respectively. Types of lines
are same as in Figure~\ref{fig:cross-angular}.
 }
 \label{fig:sn}
\end{figure}

We then calculate the S/N ratio for the cross-correlation for the
fiducial reionization case ($\zeta=40$ and $T_{\rm vir}=10^4$).  The
left panel of Fig.~\ref{fig:sn} shows the dependence of the S/N ratio on
the values of $f_{\rm NL}$.  
As discussed above, the amplification due to $f_{\rm NL}$ arises on
large scales. However, the cosmic variance is significant on such scales
Accordingly, the S/N ratio is suppressed on small $\ell_{\rm max}$.  As
$\ell_{\rm max}$ increases, the S/N ratio also goes up until $\ell \sim
100$ where the noise power spectrum of the 21~cm observation dominate
the cross-correlation signal completely. The existence of
$f_{\rm NL}$ brings relatively high S/N ratio, compared with the case
for $f_{\rm NL}=0$.  However, since the amplification due to $f_{\rm
NL}$ becomes small on small scales, the S/N ratio has weak dependence on
$f_{\rm NL}$.

Fig.~\ref{fig:sn} shows that, while the amplitude of the
cross-correlation enhances with $f_{\rm NL}$ increasing, the S/N ratio
decreases with $f_{\rm NL}$ increasing.  For example, the S/N ratio is
$2.4$ for $f_{\rm NL} =100$ and is $2.8$ for $f_{\rm NL} =10$ with
$l_{\rm max}=200$.  This is because large $f_{\rm NL}$ also increases
$C_\ell ^{21} $ appearing in the denominator in
Eq.~(\ref{eq:SNratio}). However the amplification due to $f_{\rm NL}$,
$R_{\rm NG}$, for the auto-correlation in Fig.~\ref{fig:21cm-angular} is
suppressed for small $f_{\rm NL}$, compared with $R_{\rm NG}$ for the
cross-correlation in Fig.~\ref{fig:cross-angular}.  As a result, small
values of $f_{\rm NL}$ gives the large S/N ratio.  In the case for
$f_{\rm NL} <10$, the amplification of the cross-correlation due to
$f_{\rm NL}$ becomes small. Therefore the S/N ratio also starts to
decrease with $f_{\rm NL}$ getting lower for $f_{\rm NL} < 10$.  

On the other hand, the auto power spectrum completely dominate the noise
power spectrum of Omniscope as in Fig.~\ref{fig:21cm-angular}.
Therefore the S/N ratio is large and $S/N \sim \ell_{\rm max}$ even for
$\ell _{\rm max} \sim 10$.
the Omniscope auto power-spectrum S/N ratio is large and
For comparison, we evaluate
the $S/N$ ratio in SKA, where we adopt $N=1400$, $A_{\rm eff}=45~ {\rm
m}^2$, $D=0.01~$Km and $f_{sky}=0.0056$ \citep{2011arXiv1105.1773J}. The
S/N ratio is 3.8 for $\ell_{max} =50$ for each $f_{\rm NL}$, although the
S/N ratio is 1.6 for $\ell_{max} =20$.
This suggests that in the
absence of significant foregrounds and systematics, the
auto-correlations of 21~cm is a better probe than the
cross-correlations (as expected since it depends on $b^2$), while the
cross-correlations has only 1 factor of $b$. 
Nevertheless, it is interesting to look at the cross-correlations,
since it is more likely we can rid of systematics and foregrounds that
are common to both CMB and 21~cm experiments.

The {\it rapid} reionization case with $\zeta=400$ and $T_{\rm vir}=10^5~$K
brings high S/N ratio. 
Our estimated S/N ratio for the rapid
reionization case is $2.7$ for $f_{\rm NL}=10$ and $3.2$ for $f_{\rm NL}=10$
as shown in the right panel of
Fig.~\ref{fig:sn}.
Therefore, this fact suggests that any detection of excess power in the
cross-correlation with relatively high S/N ratio implies the efficient
reionization process and the existence of high $f_{\rm, NL}$.

\section{Conclusion and Discussion}

In this paper we have studied the potential of the cross-correlation
between CMB temperature anisotropies and 21~cm fluctuations from EoR to
constrain the primordial non-Gaussianities.  Assuming the analytic
reionization model, we have utilized the effect of primordial
non-Gaussianity on the bias of the ionized fraction fluctuations.  We have
calculated the cross-correlation to the linear order and shown the
angular cross-power spectrum, while contrasting against  21~cm auto power-spectrum.

Due to the scale dependent nature of the effect of primordial
non-Gaussianity, the effect is larger at large scales. The higher
$f_{\rm NL}$ become, the more the angular power spectrum is enhanced,
and the enhancement is more significant in lower multipoles. Since the
amplitude of the cross-correlation depends on the efficiency of the
reionization, we also investigated the effect of different reionization
models on the cross power-spectrum. The overall amplitude of cross-power
in the {\it rapid} reionization model is higher than the overall
amplitude of cross-power in the slow reionization model. The
amplification due to the non-Gaussianity, $R_{\rm NG}$,  depends on the
critical density, $\delta_B$, of the
ionized bubbles. However, in our reionization model, we found that the
dependence of $\delta_B$ on the ionization parameter is weak. As a
result, $R_{\rm NG}$ are almost same in both the fiducial and {\it
rapid} reionization models. This suggests that the determination of
$f_{\rm NL}$ from $R_{\rm NG}$ does not degenerate with other reionization parameters strongly.

The degree of the amplification due to $f_{\rm NL}$ in the
cross-correlation is larger on $\ell \gtrsim 10$ than the
corresponding scale in 
auto-correlation of the 21~cm fluctuations.  However, the CMB Doppler
signal becomes small on small scales and is dominated by the primordial
CMB signal. This makes the noise large and the detection of the
cross-correlation difficult, when compared against the auto-correlation of the
21~cm fluctuations.

To access the detectability, we have calculated the signal-to-noise
(S/N) ratio of both auto- and cross-power of 
 Omniscope.  In the case of the fiducial (slow)  reionization model, the S/N ratio
is 2.4 for $f_{\rm NL}=100$ and become 2.8 for $f_{\rm NL} =10$.
Since high $f_{\rm NL}$ enhances the auto-power spectrum which increases
the noise for the cross-correlation signal, high $f_{\rm NL}$ brings
small S/N ratio and we obtain the maximum S/N ratio at $f_{\rm NL}=10$.
In the {\it rapid} reionization, the S/N ratio is enhanced for all $f_{\rm NL} >0$.
Since the amplification due to $f_{\rm NL}$, $R_{\rm NG}$, does not depends on
the reionization parameters, the enhancement of the S/N ratio in the
{\it rapid} suggest that $f_{\rm NL}$ is well determined in the {\it
rapid} reionization model.

In comparison, the S/N ratio for the auto-correlation is quite large.
Even for SKA, the S/N ratio becomes 3.8 for $l_{\rm max} =50$.
This suggests that in the
absence of significant foregrounds and systematics, the
auto-correlations of 21~cm is a better probe than the
cross-correlations (as expected since it depends on $b^2$), while the
cross-correlations contains only 1 factor of $b$. 
Nevertheless, it is interesting to look at the cross-correlations,
since it is more likely we can rid of systematics and foregrounds that
are common to both CMB and 21~cm experiments than completely clean
21~cm of all of the possible foregrounds and systematics in large scales.
In the calculation of the S/N ratio, we ignore the foreground
contamination of the cross-correlation between CMB and 21cm
fluctuations.  In reality, some of the foregrounds for 21-cm
observations also have the correlation with CMB observations
\citep{2008MNRAS.384..291A}.  This may affect significantly 
the detection of the signal.  Therefore a better model of the
foreground is essential for any 21~cm constraint on the non-Gaussianity.  The
tidal approach suggested by \citet{2012arXiv1202.5804P} can  be a
potential technique to reduce such foreground contamination from 21 cm
mapping.

In the paper, we consider only one redshift slice for the 21~cm
observation in this paper.  We can observe many redshift slices by
choosing the observation frequency.  According to
Eq.~(\ref{eq:SNratio}), taking many redshift slices would increase S/N
ratio for each $f_{\rm NL}$. As a result, multi-frequency observation of
21~cm fluctuations can bring the better constraint on $f_{\rm NL}$ than
the one redshift slice such as considered in this paper.  In particular,
the S/N ratio for the auto-correlation receives benefit richly from
multi-frequency observation. It can be expected that even the S/N ratio
for SKA become enough to measure the non-Gaussianity as studied in
\citet{2011arXiv1105.1773J}.  On the other hand, the signal of the
cross-correlation in the redshift evolution reaches a peak during the
epoch when the ionized fraction becomes a half
\citep{2006ApJ...647..840A}.  In particular, in contrast to the case of
the rapid reionization, there is possibility to utilize many redshift
slices for 21~cm fluctuations in the case of the slow reionization, as
the cross-correlation signals arises during a long period due to the
slow evolution of the ionized fractions.

Finally, we focus the signals from the EoR on large
scales. However it is well-known that the distribution of the bubbles
affect CMB anisotropies and 21~cm fluctuations
\citep{2006NewAR..50..909I, 2004ApJ...613...16F}. Therefore, the effect
of non-Gaussianity can be expected to arise on small scales. We will
leave this to future work. 
%

\section*{Acknowledgements}
We thank N. Sugiyma, M. McQuinn and U.-L. Pen for their insightful
comments. S.H. would like to acknowledge the Department of Energy
Lawrence Berkeley National Laboratory Chamberlain and Seaborg Fellowship
which supports the production of this work.

\end{document}